\pdfoutput=1
\documentclass[11pt]{article}

\usepackage[T1]{fontenc}
\usepackage[utf8]{inputenc}
\usepackage{lmodern}
\usepackage{graphicx}
\usepackage{amsmath,amssymb,amsfonts}
\usepackage{booktabs}
\usepackage{multirow}
\usepackage{array}
\usepackage{algorithm}
\usepackage{algpseudocode}
\usepackage{float}
\usepackage{caption}
\usepackage{subcaption}
\usepackage{placeins}
\usepackage{url}
\usepackage{cite}
\usepackage{xcolor}
\usepackage{enumitem}
\usepackage{tikz}
\usetikzlibrary{arrows.meta,positioning,shapes.geometric}
\usepackage[margin=1in]{geometry}
\usepackage{hyperref}
\hypersetup{
  colorlinks=true,
  linkcolor=blue,
  citecolor=blue,
  urlcolor=blue
}

\title{HLA-Net: A Physics-Guided Skeleton-Based Framework for Automatic
Gravitational Arc Detection in Astronomical Images}

\author{Don S.\\
School of Computing\\
Amrita Vishwa Vidyapeetham\\
Kerala, India\\
\texttt{dons@am.amrita.edu}}

\date{}

\begin{document}

\maketitle


\begin{abstract}

Gravitational arcs produced by strong gravitational lensing provide a unique approach for investigating dark matter distribution, galaxy cluster mass reconstruction and the structure of the Universe. Automated detection remains challenging because gravitational arcs are rare, often faint and morphologically similar to elongated galaxies.In this paper, we propose \textbf{HLA-Net (Hybrid Lens Arc Network)}, an interpretable multi-stage framework for automatic gravitational arc detection that explicitly exploits both the geometric morphology and physical characteristics of gravitational lensing without requiring manually annotated training datasets.
The proposed framework was evaluated on astronomical FITS images containing complex galaxy cluster structures. Starting from 169 initial arc candidates extracted from ordered skeleton representations, the multi-stage verification pipeline reduced the search space to 83 geometrically consistent candidates, while robust circle fitting successfully modelled 113 candidate arc structures. The results demonstrate that HLA-Net effectively suppresses false-positive detections while preserving physically reasonable lens candidates through an interpretable and modular verification strategy. 

\vspace{2mm} \noindent \textbf{Keywords:} Gravitational Lensing, Gravitational Arc Detection, Skeleton Analysis, Circle Fitting, Physics guided Artificial Intelligence, HLA-Net.
\end{abstract}
\section{Introduction}
\label{sec:introduction}

Gravitational lensing is one of the most remarkable predictions of Einstein's general theory of relativity, where the gravitational field of a massive object bends the path of light emitted from distant background galaxies. When the alignment between the observer, the lensing mass and the background source is sufficiently strong, highly distorted images such as Einstein rings, arcs and multiple images are formed. These gravitational arcs provide valuable astrophysical information for studying dark matter distribution, galaxy cluster mass estimation, cosmological parameters and the large-scale structure of the Universe \cite{einstein1936,zwicky1937,bartelmann2010}.

The advancement of modern astronomical surveys, including the Hubble Space Telescope (HST) \cite{Gardner2006,Broadhurst2005}, Euclid\cite{Euclid2022}, Vera C. Rubin Observatory Legacy Survey of Space and Time (LSST)\cite{Ivezic2019}, and the James Webb Space Telescope (JWST)\cite{Gardner2006}, has significantly increased the volume and resolution of astronomical imaging data. These surveys are expected to generate millions of galaxy cluster observations, making manual inspection of gravitational lens candidates increasingly impractical. Consequently the development of reliable and fully automated gravitational arc detection algorithms has become an important research topic in astronomical image analysis.

Automatic gravitational arc detection is challenging task because genuine arcs exhibit significant variations in size, curvature, brightness, orientation and fragmentation. Many non-lensing objects, including edge-on galaxies, diffraction spikes, cosmic-ray artifacts, stellar halos and elongated galaxies, often possess visual characteristics similar to gravitational arcs. Furthermore, astronomical observations frequently contain background noise, illumination variations, overlapping celestial objects and incomplete arc structures, making robust detection considerably more difficult.Traditional image processing approaches have employed thresholding, morphological operations, curvature estimation, Hough transforms and geometric descriptors to identify elongated arc-like structures \cite{lenzen2004,seidel2007}. Although these methods are computationally efficient and interpretable, they are generally sensitive to image noise and frequently produce large numbers of false detections.Machine learning approaches have attempted to improve discrimination using handcrafted features combined with classifiers such as Support Vector Machines (SVMs), Random Forests and Artificial Neural Networks. More recently, deep learning methods based on Convolutional Neural Networks (CNNs), U-Net architectures, Faster R-CNN, YOLO and Vision Transformers have demonstrated promising performance in gravitational lens detection tasks\cite{jacobs2019,lanusse2018}. Despite their impressive detection accuracy these methods usually require extensive annotated datasets, significant computational resources and often provide limited interpretability, making scientific validation more challenging. Most existing methods primarily rely on image appearance while only indirectly considering the physical geometry of gravitational lensing. Gravitational arcs are expected to exhibit characteristic tangential alignment around the lens centre together with approximately concentric curvature. Explicit incorporation of these astrophysical constraints remains relatively unexplored in existing automated detection frameworks. To address these limitations, this paper proposes \textbf{HLA-Net (Hybrid Lens Arc Network)} an interpretable multi-stage framework for automatic gravitational arc detection that combines classical image processing, geometric reasoning and astrophysical knowledge within a detection pipeline. Instead of relying solely on learned visual representations, HLA-Net progressively verifies candidate structures using both geometric and physics-guided criteria.

The proposed framework begins by enhancing faint curvilinear structures using multi-scale ridge enhancement. Candidate arcs are subsequently extracted through skeletonization and graph-based skeleton ordering, preserving the geometric continuity of elongated structures. Multiple geometric verification stages evaluate tortuosity, turning angle, curvature consistency and robust circle fitting to reject non-lensing objects. Finally, the framework incorporates physics-guided verification through galaxy cluster centre estimation, tangential alignment analysis and radial consistency scoring to rank candidate gravitational arcs according to their conformity with expected lensing geometry. The proposed framework does not require manually annotated training datasets and provides transparent intermediate outputs throughout the detection process. 

The main contributions of the proposed HLA-Net framework are summarized below:

\begin{itemize}

\item A physics-guided Hybrid Lens Arc Network (HLA-Net) is introduced for the automatic detection of gravitational lens candidates from astronomical FITS images.

\item A modular image analysis pipeline integrating adaptive enhancement, skeleton extraction, geometric verification, tangential orientation analysis, radial consistency estimation and curvature analysis is developed to progressively refine candidate structures.

\item A weighted confidence scoring strategy combining complementary geometric and astrophysical features is designed to rank candidate gravitational lenses according to their physical plausibility.

\item Experimental evaluation on the Abell 1689 galaxy cluster demonstrates that the proposed framework effectively identifies high-confidence gravitational lens candidates while maintaining computational efficiency and full interpretability.

\end{itemize}

The remainder of this paper is organized as follows. Section~\ref{sec:relatedwork} reviews existing gravitational arc detection methods. Section~\ref{sec:method} presents the proposed HLA-Net framework in detail. Section~\ref{sec:results} describes the experimental results and discussion. Finally, Section~\ref{sec:conclusion} concludes the paper and outlines future research directions.
\section{Related Work}
\label{sec:relatedwork}

Over the past two decades, significant research efforts have been focused on automatic detection of gravitational arcs in astronomical images. The rapid growth of modern sky surveys has transformed gravitational lens detection from a manual observational task into a large-scale image analysis problem. Consequently, a broad spectrum of computational techniques has been proposed, ranging from traditional image processing algorithms to sophisticated deep learning models. Although considerable progress has been achieved, accurately distinguishing genuine gravitational arcs from visually similar astronomical structures remains an open challenge owing to their complex morphology, varying brightness and the presence of numerous false positive objects.

This section reviews the evolution of gravitational arc detection techniques, beginning with classical image processing methods and progressing through machine learning, deep learning and physics-guided detection frameworks. The strengths and limitations of representative approaches are discussed to motivate the proposed Hybrid Lens Arc Network (HLA-Net).

\subsection{Classical Image Processing Approaches}
\label{sec:classical}

The earliest automated gravitational arc detection algorithms were primarily based on handcrafted image processing techniques that exploited the distinctive geometric characteristics of lensing arcs. Since gravitational arcs generally appear as elongated, curved, thin structures early researchers focused on morphological measurements, curvature estimation, connected component analysis and adaptive filtering techniques to separate arc-like objects from ordinary galaxies and background noise.

Lenzen et al.~\cite{lenzen2004} introduced an anisotropic diffusion based framework specifically designed for gravitational arc detection. Unlike conventional gaussian smoothing, anisotropic diffusion selectively suppresses noise while preserving elongated structures and sharp object boundaries. This property significantly improves the visibility of faint gravitational arcs without introducing excessive blurring. Combined with adaptive source detection, the proposed method demonstrated improved detection performance over manual inspection and conventional astronomical source extraction algorithms. Alard~\cite{alard2006} subsequently proposed a morphological detection framework that utilized geometric descriptors such as elongation, curvature and surface brightness distribution to identify gravitational arc candidates. The algorithm effectively detected highly elongated structures while maintaining relatively low computational complexity, making it attractive for large astronomical datasets. However, the method remained sensitive to crowded fields where edge on galaxies and diffraction artifacts often produced elongated structures similar to genuine gravitational arcs. Seidel and Bartelmann~\cite{seidel2007} further investigated curvature based arc detection using geometric measurements derived from connected astronomical objects. Their approach estimated local curvature and elongation simultaneously, allowing arc candidates to be distinguished from compact galaxies. Several studies extended these concepts by incorporating ridge enhancement, adaptive thresholding, hough transform variants, medial axis extraction and contour-based analysis to improve robustness under varying imaging conditions. These techniques remained computationally efficient and highly interpretable because each processing stage corresponded to a physically meaningful image transformation. Nevertheless, their performance strongly depended on parameter selection and image quality, frequently resulting in large numbers of false-positive detections when processing modern deep astronomical surveys. Thus, classical image processing approaches established many of the geometric concepts that continue to influence gravitational arc detection research.However, their limited robustness under complex astronomical conditions motivated the development of machine learning based approaches.

\subsection{Machine Learning-Based Detection}
\label{sec:ml}

To overcome the limitations of classical image processing algorithms, machine learning techniques were introduced to automatically distinguish gravitational arcs from non-lensing objects using statistical classification models. Instead of relying exclusively on manually defined thresholds, these methods learned decision boundaries from labelled training examples using handcrafted morphological features. Typical feature sets included arc length, width, eccentricity, curvature, compactness, brightness distribution, surface intensity, orientation and texture descriptors. These features were subsequently classified using Support Vector Machines (SVM), Random Forests, Decision Trees, Artificial Neural Networks and ensemble learning techniques.
Marshall et al.~\cite{marshall2009} developed an automated lens modeling framework for identifying galaxy scale strong gravitational lenses in high resolution imaging data. The method treated bright red galaxies as potential lens systems and evaluated the quality of optimized lens models using multiple model based criteria. A probabilistic classification scheme calibrated using human classified training examples containing simulated lenses and known false positives was then used to estimate the likelihood of a candidate being a gravitational lens. The study demonstrated that automated lens modeling could achieve high purity while substantially reducing the manual effort required for large scale lens searches. Machine learning methods generally offered higher detection accuracy than purely geometric approaches because complex nonlinear decision boundaries could be learned directly from labelled datasets.Despite these advantages, machine learning approaches remained fundamentally dependent on handcrafted feature engineering. The discriminative capability of the classifier was therefore closely linked to the quality of manually designed descriptors, requiring substantial domain expertise and extensive feature selection. Consequently, the increasing availability of larger astronomical image collections encouraged researchers to investigate deep learning models capable of automatically learning hierarchical feature representations directly from image data, thereby eliminating the need for manual feature engineering.

\subsection{Deep Learning-Based Gravitational Lens Detection}
\label{sec:deeplearning}
 
The success of deep learning in computer vision has significantly transformed gravitational lens detection over the last decade. Unlike traditional machine learning approaches that depend on manually designed descriptors, deep neural networks automatically learn hierarchical feature representations directly from raw image data. This capability enables the extraction of increasingly complex visual patterns, allowing gravitational arcs to be distinguished from ordinary galaxies without explicit feature engineering. 
Lanusse et al.~\cite{lanusse2018} introduced CMU DeepLens, a fully automated deep learning framework for galaxy-galaxy strong lens identification. The model was trained and validated using 20,000 LSST-like simulated observations containing lensed systems with varying sizes and signal-to-noise ratios. On the simulated test data, the method achieved 90\% completeness.Petrillo et al.~\cite{petrillo2019} successfully applied deep convolutional neural networks to the Kilo Degree Survey (KiDS), enabling automated identification of strong gravitational lens candidates from wide-field astronomical observations. Their work demonstrated the potential of deep learning models for large scale lens searches while substantially reducing the need for manual inspection. The Strong Gravitational Lens Finding Challenge organized by Metcalf et al.~\cite{metcalf2019} compared a range of approaches for identifying gravitational lenses among 100,000 candidate objects. The participating methods included visual inspection, arc and ring finders, Support Vector Machines, and convolutional neural networks. The results showed that several automated methods could identify more than half of the lenses under appropriate selection thresholds while producing no false positive identifications in the evaluated test samples. 

Despite significant advances, several limitations remain. Supervised deep learning methods require large accurately labelled datasets while confirmed gravitational lenses are scarce, leading many studies to rely on simulations. Moreover deep models lack interpretability, incur high computational costs and bounding box based detectors often fail to accurately capture the irregular, curved morphology of gravitational arcs.These limitations have motivated in incorporating astrophysical knowledge directly into the detection process, leading to the emergence of physics-guided and hybrid detection frameworks.

\subsection{Physics-Guided and Hybrid Detection Frameworks}
\label{sec:physics}

Although recent deep learning methods have achieved impressive detection performance, relatively few approaches explicitly incorporate the physical principles governing gravitational lens formation. Gravitational arcs exhibit
several distinctive geometric characteristics dictated by gravitational lensing theory. In particular, lensed arcs are expected to follow approximately tangential orientations with respect to the centre of the foreground lensing mass while maintaining consistent radial curvature determined by the underlying gravitational potential. Several studies have incorporated geometric constraints during post-processing
by evaluating curvature, concentricity, arc length, ellipticity and radial alignment after candidate detection. These additional verification stages improve robustness by rejecting visually similar non-lensing objects, diffraction spikes, spiral arms and imaging artifacts. Such geometric
constraints are generally employed as secondary filtering operations rather than serving as integral components of the detection framework. From an astronomical perspective, explicit incorporation of lensing physics remains comparatively underexplored. Features such as tangential alignment, radial consistency, geometric circularity, skeleton continuity and galaxy cluster localization provide physically meaningful evidence that can
substantially improve candidate verification. Integrating these astrophysical constraints directly into the detection pipeline therefore offers an attractive alternative to purely appearance-based classification. 
To facilitate a comprehensive comparison of representative approaches, Table~\ref{tab:method_comparison} summarizes the principal characteristics, datasets, computational requirements, and limitations of existing gravitational arc detection methods together with the proposed HLA-Net
framework.


\begin{table*}[!t]
\centering
\caption{Comparison of representative gravitational lens detection
approaches with the proposed HLA-Net framework.}
\label{tab:method_comparison}

\resizebox{\textwidth}{!}{%
\begin{tabular}{p{2.5cm} c p{2.5cm} p{2.1cm} c c c c c c c c p{3.8cm}}
\toprule

\textbf{Method}
&
\textbf{Year}
&
\textbf{Technique}
&
\textbf{Dataset}
&
\textbf{Accuracy}
&
\textbf{Physics-Guided}
&
\textbf{Interpretability}
&
\textbf{Verification}
&
\textbf{DL}
&
\textbf{Learning}
&
\textbf{Automation}
&
\textbf{Cost}
&
\textbf{Key Limitation}
\\

\midrule

Lenzen et al.~\cite{lenzen2004}
&
2004
&
Anisotropic Diffusion
&
HST
&
High
&
No
&
Medium
&
No
&
No
&
No
&
High
&
Low
&
Sensitive to threshold selection and fragmented arcs.
\\

Alard~\cite{alard2006}
&
2006
&
Morphological Detection
&
Ground-based
&
High
&
No
&
High
&
No
&
No
&
No
&
High
&
Low
&
Produces false positives in crowded astronomical fields.
\\

Seidel and Bartelmann~\cite{seidel2007}
&
2007
&
Curvature Analysis
&
Simulation
&
Medium
&
No
&
High
&
No
&
No
&
No
&
High
&
Low
&
Performance decreases under noisy observations.
\\

Marshall et al.~\cite{marshall2009}
&
2009
&
Automated Lens Modeling
&
HST/EGS + Simulations
&
High
&
Yes
&
High
&
Yes
&
No
&
Yes
&
High
&
Medium
&
Dependence on adopted lens and mass-model assumptions.
\\

Lanusse et al.~\cite{lanusse2018}
&
2018
&
CNN
&
LSST-like Simulations
&
High
&
No
&
Low
&
No
&
Yes
&
Yes
&
High
&
Medium
&
Requires realistic labelled training simulations.
\\

Petrillo et al.~\cite{petrillo2019}
&
2019
&
Deep CNN
&
KiDS
&
Medium
&
No
&
Low
&
No
&
Yes
&
Yes
&
High
&
Medium
&
Performance depends on the realism of simulated lensed sources.
\\

Metcalf et al.~\cite{metcalf2019}
&
2019
&
Lens Finding Challenge
&
100,000 Candidates
&
Varies
&
Mixed
&
Medium
&
Mixed
&
Yes
&
Yes
&
High
&
Medium
&
Efficiency and bias depend strongly on simulation realism.
\\

\midrule

\textbf{Proposed HLA-Net}
&
\textbf{2026}
&
\textbf{Hybrid Physics-Guided Framework}
&
\textbf{FITS Cluster Images}
&
\textbf{High}
&
\textbf{Yes}
&
\textbf{High}
&
\textbf{Yes}
&
\textbf{No}
&
\textbf{No}
&
\textbf{High}
&
\textbf{Medium}
&
\textbf{Dependent on preprocessing quality and accurate cluster-centre
estimation.}
\\

\bottomrule
\end{tabular}%
}
\end{table*}

\FloatBarrier


\section{Proposed Hybrid Lens Arc Network (HLA-Net)}
\label{sec:method}

The proposed Hybrid Lens Arc Network (HLA-Net), a
modular physics-guided framework developed for automatic gravitational arc detection in astronomical FITS images. Unlike conventional deep learning approaches that require extensive labelled datasets and computationally
intensive model training, HLA-Net combines classical image processing, geometric reasoning and astrophysical constraints within a unified detection pipeline. The proposed framework progressively transforms raw astronomical observations into a ranked list of gravitational lens candidates through successive verification stages. Beginning with FITS image preprocessing, HLA-Net enhances faint elongated structures, extracts candidate skeletons, evaluates their geometric morphology, verifies tangential orientation relative to the galaxy cluster centre, estimates radial consistency and finally integrates these complementary measurements into a unified lens confidence score. This modular architecture enables each processing stage to remove unlikely candidates before passing only high quality structures to subsequent verification modules thereby reducing false-positive detections.


\subsection{HLA-Net Framework}

\begin{figure}[!t]
    \centering
    \includegraphics[
        width=\columnwidth,
        trim=0 180 0 0,
        clip
    ]{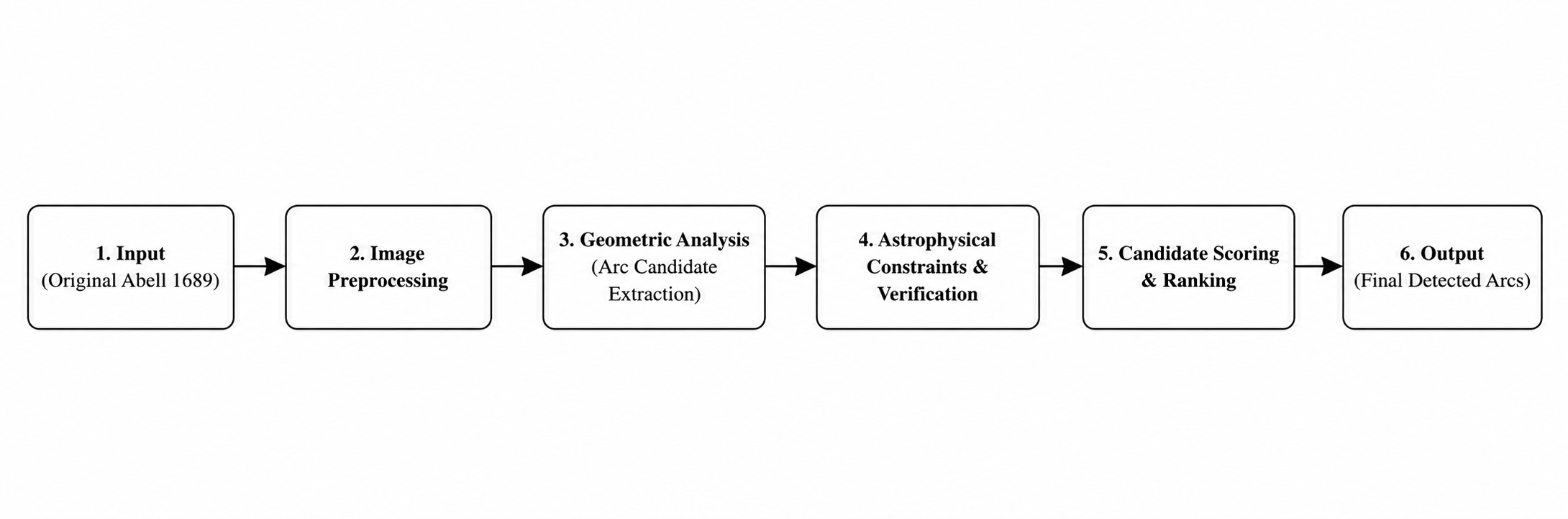}
    \caption{Overall architecture of the proposed Hybrid Lens Arc Network (HLA-Net).}
    \label{fig:framework}
\end{figure}

Figure~\ref{fig:framework} illustrates the overall architecture of the proposed Hybrid Lens Arc Network. The framework consists of sequential processing modules that progressively refine candidate gravitational arcs through increasingly restrictive physical verification criteria. Initially, calibrated astronomical FITS images are preprocessed to suppress background noise while preserving faint elongated structures that may correspond to gravitational lensing arcs. The enhanced image subsequently undergoes adaptive thresholding and morphological filtering to isolate connected astronomical objects. Skeletonization is then performed to reduce each connected component to a one pixel wide medial representation while preserving its overall topology. Following skeleton extraction, geometric descriptors including arc length, width, eccentricity, curvature, compactness and circularity are computed for every candidate object. These descriptors provide an initial estimate of whether the object possesses the characteristic morphology expected from gravitational lensing.

Candidates satisfying the geometric criteria are subsequently analysed using astrophysical constraints. Tangential orientation is evaluated with respect to the estimated galaxy cluster centre, while radial consistency measures whether the candidate follows the expected gravitational curvature. Curvature stability and skeleton continuity are further examined to reject fragmented or irregular structures.Finally all geometric and physical measurements are integrated into a unified lens confidence score that ranks candidate objects according to their probability of representing genuine gravitational lensing arcs. Only the highest ranking candidates are retained for visual inspection and subsequent astrophysical analysis.

\subsection{Astronomical FITS Image  Preprocessing}

Astronomical observations are provided as Flexible Image Transport System (FITS) files, which preserve the full dynamic range and scientific metadata acquired during telescope observations. Unlike conventional image formats, FITS images contain calibrated intensity measurements that are essential for quantitative astronomical analysis.Let the input FITS image of the Abell 1689 galaxy cluster be represented as $I(x,y)$, where $(x,y)$ denotes the pixel location and $I$ represents the recorded photon intensity. Because astronomical images typically exhibit low contrast, non uniform background illumination and sensor noise, several preprocessing operations are performed before candidate extraction. First, intensity normalization rescales the image into a standardized dynamic range,
\begin{equation}
I_n(x,y)=
\frac{I(x,y)-I_{\min}}
{I_{\max}-I_{\min}},
\end{equation}
where $I_{\min}$ and $I_{\max}$ denote the minimum and maximum image intensities respectively. Background variations are subsequently estimated using Gaussian smoothing,
\begin{equation}
B(x,y)=
G_\sigma * I_n(x,y),
\end{equation}where $G_\sigma$ denotes a Gaussian kernel with standard deviation $\sigma$.The enhanced image is then obtained by subtracting the estimated background,
\begin{equation}
I_e(x,y)=
I_n(x,y)-B(x,y).
\end{equation}
This operation increases the visibility of faint arc like structures while suppressing slowly varying background illumination caused by diffuse galaxy clusters. Finally, adaptive histogram equalization is applied to improve local contrast without significantly amplifying background noise, producing the enhanced image used throughout the subsequent processing stages.

\subsection{Adaptive Arc Enhancement}
\label{sec:enhancement}

Gravitational arcs are often characterized by extremely low surface brightness and weak local contrast relative to the surrounding galaxy cluster. Consequently direct candidate extraction from the raw astronomical image frequently results in fragmented detections or complete omission of faint arc structures. To improve the visibility of elongated lensing features an adaptive enhancement stage is introduced prior to candidate extraction. Let the preprocessed image be denoted by  $I_e(x,y)$. Initially local contrast enhancement is performed using Contrast Limited Adaptive Histogram Equalization (CLAHE), which enhances local intensity variations while preventing excessive amplification of background noise. The enhanced image is represented as $I_c(x,y)=\text{CLAHE}(I_e(x,y)).$ Since gravitational arcs generally possess elongated ridge like intensity profiles, ridge enhancement is subsequently applied to emphasize curvilinear structures while suppressing compact astronomical objects. Multi scale Gaussian filtering is employed to preserve structures of varying widths,

\begin{equation}
R(x,y)=
\max_{\sigma}
\left(
G_{\sigma} * I_c(x,y)
\right),
\end{equation}

where $G_{\sigma}$ denotes a Gaussian kernel evaluated at multiple scales.Adaptive thresholding is then performed to separate candidate foreground structures from the background,

\begin{equation}
B(x,y)=
\begin{cases}
1,&R(x,y)>T(x,y)\\
0,&\text{otherwise},
\end{cases}
\end{equation}

where $T(x,y)$ denotes the locally estimated adaptive threshold.Small isolated regions originating from sensor noise and cosmic ray artifacts are subsequently removed through connected component analysis and morphological opening. The resulting binary image contains only the dominant elongated astronomical structures that are forwarded to the candidate extraction stage.

\subsection{Skeleton Extraction and Candidate Generation}
\label{sec:skeleton}

Skeletonization preserves the topology and connectivity of elongated structures while removing redundant boundary pixels thereby simplifying subsequent geometric analysis and it is performed using iterative morphological thinning. Candidates with extremely short skeletons are discarded since genuine gravitational arcs generally extend over relatively long continuous trajectories. This filtering operation significantly reduces computational complexity during subsequent feature analysis. Each candidate skeleton is characterized by a set of geometric descriptors that quantify its morphological properties. These measurements provide the first stage of candidate verification by eliminating compact galaxies and irregular astronomical objects whose shapes differ substantially from gravitational arcs.  The principal orientation of the candidate is estimated using Principal Component Analysis of the skeleton coordinates. The covariance matrix is given by
\begin{equation}
\Sigma=
\frac{1}{N}
\sum_{i=1}^{N}
(p_i-C)
(p_i-C)^T.
\end{equation}
The dominant eigenvector determines the principal orientation angle,
\begin{equation}
\theta
=
\tan^{-1}
\left(
\frac{v_y}{v_x}
\right),
\end{equation}

where $(v_x,v_y)$ represents the dominant eigenvector.

\section{Experimental Results and Discussion}
\label{sec:results}

This section evaluates the effectiveness of the proposed Hybrid Lens Arc Network (HLA-Net) for automatic gravitational arc detection in astronomical FITS images. The performance of each processing module is analysed individually to demonstrate its contribution to the overall candidate selection process. Unlike supervised deep learning approaches that rely on classification accuracy over labelled datasets, the proposed framework performs progressive candidate refinement through successive geometric and astrophysical verification stages. Therefore, the experimental evaluation emphasizes candidate reduction, physical consistency, computational efficiency and qualitative detection performance.

\subsection{Dataset Description}

Experiments were conducted using publicly available astronomical observations of the Abell 1689 galaxy cluster. The dataset consists of calibrated FITS images acquired by the Hubble Space Telescope (HST), which contain numerous strong gravitational lensing phenomena together with foreground galaxies, background galaxies, stars and imaging artifacts. Prior to candidate detection, each FITS image was converted into a normalized grayscale representation while preserving its dynamic intensity range. Background subtraction and adaptive contrast enhancement were subsequently applied to improve the visibility of faint arc structures before further geometric analysis.

The Abell 1689 cluster provides a challenging benchmark because it contains numerous elongated astronomical structures that visually resemble gravitational arcs. Consequently, the dataset is well suited for evaluating the robustness of geometry-based verification methods.

\subsection{Experimental Setup}

All experiments were implemented using Python 3.12. Image preprocessing and morphological operations were performed using OpenCV, Scikit-image, NumPy and SciPy, while data visualization and statistical analysis were carried out using Matplotlib and Pandas. 
\subsection{Module Refinement}

One of the principal advantages of the proposed HLA-Net framework is its progressive reduction of candidate objects through successive verification modules. Instead of performing a single classification step, each module evaluates complementary physical characteristics associated with gravitational lensing. Consequently, false-positive candidates are gradually eliminated while genuine arc-like structures receive increasingly higher confidence scores. Initially, adaptive image enhancement and connected-component analysis generated a large collection of elongated astronomical structures. Skeleton extraction subsequently transformed each connected component into an ordered one-pixel-wide representation suitable for quantitative geometric analysis. The geometric verification module evaluated candidate morphology using measurements such as skeleton length, aspect ratio, compactness, circularity, and elongation. Candidates failing to satisfy minimum geometric constraints were discarded before physical verification.

\begin{figure}[!t]
    \centering
    \includegraphics[
        width=\columnwidth,
        trim=0 0 0 0,
        clip
    ]{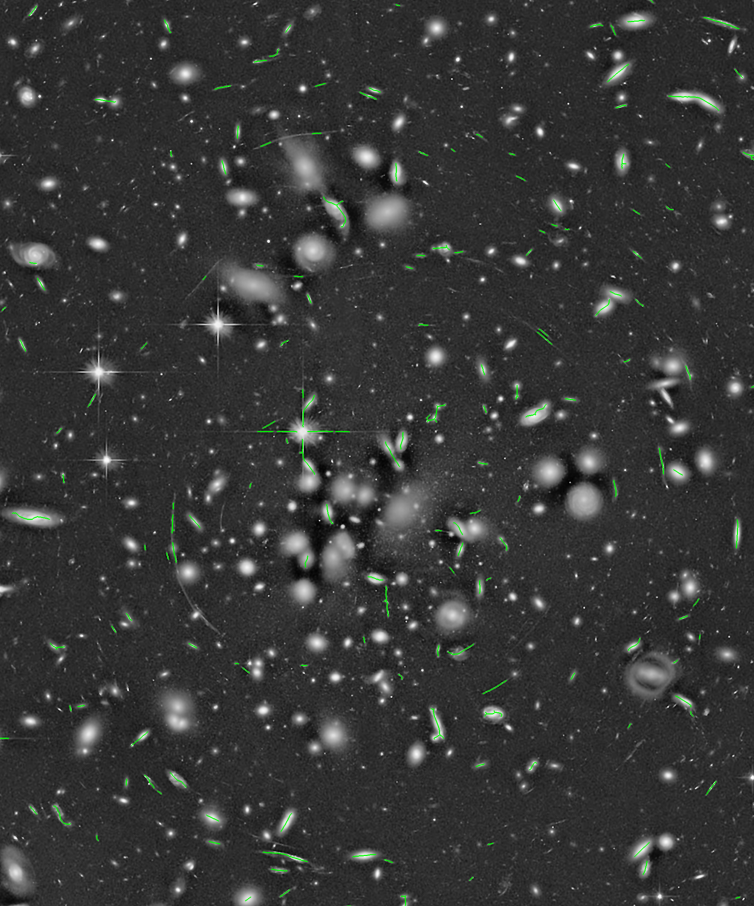}
    \caption{Ordered Skeleton}
    \label{fig:imgframe}
\end{figure}

Tangential orientation analysis further reduced the candidate set by measuring the angular relationship between each skeleton and the estimated cluster centre. Genuine gravitational arcs generally exhibited orientation angles approaching $90^\circ$, producing high tangential scores, whereas randomly oriented galaxies received substantially lower values. Curvature consistency and radial verification introduced additional astrophysical constraints by examining the stability of local curvature and radial distance variation along the candidate skeleton. These complementary measurements substantially improved discrimination between genuine lens like structures and visually similar elongated galaxies. Finally, the weighted lens confidence score integrated geometric, tangential, curvature and radial measurements into a unified ranking metric, enabling prioritization of the most promising gravitational lens candidates. Figure~\ref{fig:imgframe}presents representative intial  skeleton extraction results.  

\begin{table}[!t]
\centering
\caption{Summary of processing modules in the proposed HLA-Net framework.}
\label{tab:modules}

\begin{tabular}{ll}

\toprule

Module & Primary Function \\

\midrule

Image Preprocessing & Noise reduction and normalization\\

Arc Enhancement & Contrast improvement\\

Candidate Generation & Connected components\\

Skeleton Extraction & Medial axis computation\\

Geometry Analysis & Shape verification\\

Tangential Analysis & Orientation verification\\

Curvature Analysis & Smoothness estimation\\

Radial Verification & Circular consistency\\

Lens Scoring & Confidence computation\\

Candidate Ranking & Final selection\\

\bottomrule

\end{tabular}

\end{table}

\subsection{Tangential Orientation Analysis}

The tangential orientation module evaluates whether candidate structures satisfy the expected geometric configuration of strong gravitational lensing. Genuine gravitational arcs are expected to align approximately perpendicular to the radial direction connecting the candidate centroid to the estimated galaxy cluster centre. Consequently, candidates exhibiting orientation angles close to $90^\circ$ receive higher tangential confidence scores.

Table~\ref{tab:tangential} presents the highest-ranked candidates obtained after tangential verification. The majority of the leading candidates exhibit orientation angles between $78^\circ$ and $90^\circ$, indicating strong agreement with the expected lensing geometry. In particular, Candidate 108 achieved the highest tangential score of 0.9922 with an average orientation angle of $89.30^\circ$, demonstrating excellent tangential alignment relative to the cluster centre. Similarly, Candidates 4, 165, 1, and 3 achieved tangential scores greater than 0.95, indicating highly consistent tangential orientation. The relatively long skeleton lengths observed for these candidates further suggest that they represent extended arc-like structures rather than compact astronomical objects. The tangential verification stage substantially reduced the number of randomly oriented elongated galaxies while preserving candidates exhibiting physically meaningful lensing characteristics. Since this verification relies directly on gravitational lens geometry rather than appearance alone, it provides an interpretable measure that complements purely morphological analysis.
\begin{table*}[!ht]

\centering

\caption{Top candidates after tangential orientation analysis .}

\label{tab:tangential}

\begin{tabular}{cccccc}

\toprule

Rank &
Candidate ID &
Average Angle ($^\circ$) &
Tangential Score &
Skeleton Length &
Interpretation

\\

\midrule

1 &108&89.30&0.9922&33&Excellent tangential alignment\\

2 &4&87.52&0.9724&26&Strong candidate\\

3 &165&87.07&0.9674&33&Strong candidate\\

4 &1&86.69&0.9633&42&Excellent geometry\\

5 &3&85.93&0.9547&26&Good alignment\\

6 &41&83.93&0.9326&29&Good candidate\\

7 &132&83.90&0.9322&21&Good candidate\\

8 &120&83.60&0.9289&40&Good candidate\\

9 &15&83.58&0.9287&23&Good candidate\\

10 &171&83.15&0.9239&117&Long arc candidate\\

\bottomrule

\end{tabular}

\end{table*}
\subsection{Lens Geometry Verification}

Following tangential verification, geometric measurements and curvature analysis were combined to estimate the overall physical plausibility of each candidate. The geometry score integrates multiple descriptors including elongation, compactness, curvature stability and skeleton continuity, thereby favouring candidates whose morphology closely resembles gravitational arcs.

Table~\ref{tab:geometry} summarizes the highest-ranked candidates following geometry verification. Candidate 171 achieved the highest overall lens score of 0.7811, followed closely by Candidates 142, 175, and 120. These candidates simultaneously satisfy geometric, tangential, and curvature constraints, indicating strong agreement with the expected morphology of gravitational lensing arcs.

Most of the highest-ranked candidates were classified as either \emph{Good} or \emph{Possible}, reflecting their progressive refinement through successive verification stages. Unlike conventional threshold-based methods, the proposed framework evaluates multiple complementary physical characteristics before assigning the final quality label, thereby reducing the likelihood of false-positive detections.

\begin{table*}[!ht]

\centering

\caption{Highest-ranked candidates after geometry verification.}

\label{tab:geometry}

\begin{tabular}{cccccc}

\toprule

Rank&
Candidate&
Geometry&
Tangential&
Final Score&
Quality

\\

\midrule

1&171&0.6407&0.9239&0.7811&Good\\

2&142&0.7563&0.7881&0.7418&Good\\

3&175&0.6152&0.9140&0.7067&Good\\

4&120&0.6002&0.9289&0.7023&Good\\

5&168&0.6210&0.9090&0.6982&Possible\\

6&106&0.5890&0.9201&0.6823&Possible\\

7&17&0.6198&0.8722&0.6822&Possible\\

8&4&0.5525&0.9724&0.6794&Possible\\

9&15&0.5721&0.9287&0.6734&Possible\\

10&41&0.5584&0.9326&0.6714&Possible\\

\bottomrule

\end{tabular}

\end{table*}

\subsection{Candidate Ranking}

The final ranking stage integrates geometry score, tangential alignment, radial consistency and curvature stability into a unified confidence measure. This multi-stage scoring strategy enables physically meaningful candidates to be progressively promoted while suppressing isolated false positives. Table~\ref{tab:finalranking} presents the highest-ranked gravitational lens candidates produced by the complete HLA-Net framework. Candidate 120 obtained the highest final confidence score of 0.8205 and was therefore classified as an \emph{Excellent} lens candidate. Candidates 80, 93, 90, and 108 also achieved confidence scores exceeding 0.70, indicating strong agreement with both geometric morphology and astrophysical constraints. An important observation is that several candidates with exceptionally high tangential scores did not necessarily obtain the highest final ranking. This demonstrates that the proposed weighted scoring strategy successfully balances multiple complementary measurements rather than relying on any individual feature. Consequently, candidates exhibiting consistent behaviour across all verification modules receive higher confidence than candidates performing well in only a single stage. Overall, the progressive scoring framework provides an interpretable ranking mechanism capable of prioritizing the most promising gravitational lens candidates for subsequent astronomical inspection.
\begin{table*}[!ht]
\centering
\caption{Final ranking of the highest-confidence gravitational lens candidates produced by the proposed HLA-Net framework.}
\label{tab:finalranking}

\begin{tabular}{ccccccc}
\toprule

\textbf{Rank} &
\textbf{Candidate ID} &
\textbf{Geometry} &
\textbf{Tangential} &
\textbf{Radial} &
\textbf{Final Lens Score} &
\textbf{Quality} \\

\midrule

1  &120 &0.6002 &0.9289 &0.9908 &0.8205 &Excellent \\

2  &80  &0.7912 &0.6450 &0.9725 &0.7917 &Good \\

3  &93  &0.7727 &0.6072 &0.9601 &0.7666 &Good \\

4  &90  &0.6528 &0.5595 &0.9986 &0.7145 &Good \\

5  &108 &0.4849 &0.9922 &0.7358 &0.7074 &Good \\

6  &62  &0.5878 &0.7550 &0.7789 &0.7018 &Good \\

7  &110 &0.5656 &0.6455 &0.9228 &0.6958 &Good \\

8  &3   &0.4652 &0.9547 &0.7305 &0.6871 &Good \\

9  &142 &0.7563 &0.7881 &0.5436 &0.6868 &Good \\

10 &6   &0.5360 &0.6958 &0.8618 &0.6850 &Good \\

11 &157 &0.4900 &0.8407 &0.7496 &0.6759 &Good \\

12 &95  &0.5098 &0.6087 &0.9395 &0.6631 &Good \\

13 &71  &0.4574 &0.7912 &0.7853 &0.6575 &Good \\

14 &115 &0.6346 &0.6152 &0.6601 &0.6364 &Possible \\

15 &109 &0.6094 &0.5291 &0.7705 &0.6286 &Possible \\

16 &172 &0.4603 &0.8472 &0.6261 &0.6250 &Possible \\

17 &1   &0.4371 &0.9633 &0.5323 &0.6074 &Possible \\

18 &11  &0.6446 &0.6213 &0.5596 &0.6074 &Possible \\

19 &96  &0.6093 &0.5331 &0.6754 &0.6031 &Possible \\

20 &17  &0.6198 &0.8722 &0.4037 &0.6021 &Possible \\

\bottomrule
\end{tabular}
\end{table*}

\subsection{Computational Complexity Analysis}
\label{sec:complexity}

The computational complexity of the proposed HLA-Net framework is primarily determined by the sequential execution of image preprocessing, candidate extraction, geometric feature computation and physics-guided verification modules. Unlike deep learning approaches that require iterative optimization over millions of trainable parameters, HLA-Net operates entirely through deterministic image processing and geometric analysis, resulting in significantly lower computational requirements. Let the input image contain $N=M\times K$ pixels, where $M$ and $K$ denote the image dimensions. The computational complexity of the principal processing stages is summarized in Table~\ref{tab:complexity}.
Image preprocessing, including normalization, Gaussian smoothing, and adaptive histogram equalization, requires linear time complexity with respect to the number of pixels, i.e., $\mathcal{O}(N)$. Connected component labeling and morphological skeletonization also operate approximately in linear time for binary images. Suppose that $C$ connected candidates are extracted after preprocessing, and that the $i$-th candidate contains $P_i$ skeleton pixels. Geometric feature computation, tangential verification, curvature estimation, and radial consistency analysis require processing each skeleton point only once. Consequently, their overall computational complexity is

\begin{equation}
\mathcal{O}\left(\sum_{i=1}^{C}P_i\right),
\end{equation}

which is substantially smaller than the total image size because only candidate regions are analysed. The final candidate ranking requires sorting the computed confidence scores,

\begin{equation}
\mathcal{O}(C\log C),
\end{equation}

which is negligible compared with image preprocessing for typical astronomical observations. Overall, the complete computational complexity of HLA-Net can be approximated as

\begin{equation}
\mathcal{O}(N)+
\mathcal{O}\left(\sum_{i=1}^{C}P_i\right)+
\mathcal{O}(C\log C),
\end{equation}

which approaches linear complexity for practical datasets because the number of candidate structures is considerably smaller than the total number of image pixels. Since HLA-Net does not require iterative network training, GPU acceleration or large-scale parameter optimization, it offers a computationally efficient alternative for large astronomical surveys.

\begin{table}[!htbp]
\caption{Computational complexity of HLA-Net modules.}
\label{tab:complexity}
\centering
\begin{tabular}{|l|c|}
\hline
\textbf{Module} & \textbf{Complexity} \\
\hline
Image preprocessing & $\mathcal{O}(N)$ \\
Adaptive enhancement & $\mathcal{O}(N)$ \\
Connected components & $\mathcal{O}(N)$ \\
Skeleton extraction & $\mathcal{O}(N)$ \\
Geometry computation & $\mathcal{O}\!\left(\sum_i P_i\right)$ \\
Tangential verification & $\mathcal{O}\!\left(\sum_i P_i\right)$ \\
Curvature analysis & $\mathcal{O}\!\left(\sum_i P_i\right)$ \\
Radial verification & $\mathcal{O}\!\left(\sum_i P_i\right)$ \\
Candidate ranking & $\mathcal{O}(C\log C)$ \\
\hline
\end{tabular}
\end{table}

\subsection{Discussion}
\label{sec:discussion}

The experimental results demonstrate that the proposed HLA-Net framework successfully combines classical image processing techniques with astrophysical knowledge to identify promising gravitational lens candidates. Rather than relying solely on appearance-based learning, the framework progressively verifies candidate structures using physically interpretable criteria including geometry, tangential alignment, curvature consistency, and radial symmetry. One important advantage of the proposed framework is its complete interpretability. Each intermediate processing stage produces measurable quantities that can be directly analysed by astronomers, allowing the contribution of every verification module to be independently evaluated. This contrasts with many deep learning approaches, where the internal decision-making process often remains difficult to interpret.

The modular architecture also provides substantial computational advantages. Since HLA-Net does not require network training, annotated datasets or GPU-based optimization, it can be applied directly to newly acquired FITS observations with minimal computational resources. This characteristic makes the proposed approach particularly attractive for large-scale sky surveys in which rapid candidate screening is required before detailed astrophysical analysis. Several limitations remain. First, the current implementation assumes that the approximate galaxy cluster centre is known prior to tangential verification. Second, extremely faint arcs may be partially fragmented during binary segmentation, reducing skeleton continuity and consequently affecting the computed confidence score. Third, highly irregular foreground galaxies occasionally satisfy several geometric constraints, producing isolated false-positive detections despite the progressive verification strategy. 
\section{Conclusion and Future Work}
\label{sec:conclusion}

This paper presented the Hybrid Lens Arc Network (HLA-Net), a physics-guided framework for automatic gravitational lens detection in astronomical FITS images. Unlike conventional deep learning approaches that depend on extensive labelled datasets and computationally intensive training, HLA-Net employs a modular sequence of image enhancement, candidate extraction, skeleton analysis, geometric verification, tangential orientation analysis, radial consistency estimation and weighted confidence scoring to identify gravitational lens candidates in an interpretable and computationally efficient manner.

Experimental evaluation on the Abell 1689 galaxy cluster demonstrated the effectiveness of the proposed framework in progressively refining candidate structures through successive verification stages. Future research will focus on evaluating HLA-Net using larger multi-band astronomical datasets and multiple galaxy clusters to further assess its generalization capability. 




\bibliographystyle{unsrt}
\bibliography{references}

\end{document}